\documentclass[a4paper]{jpconf}
\usepackage{graphicx}
\usepackage{citesort}
\begin{document}
\title{Constancy of energy partition in central heavy-ion reactions at
intermediate energies}

\author{Z Basrak$^1$, Ph Eudes$^2$, M Zori\'c$^1$ and F S\'ebille$^2$}

\address{$^1$ Rud\hspace*{-0.76ex}{\raise 1.22ex\hbox{\vrule
   height 0.06ex width 0.28em}}er
   Bo\v{s}kovi\'{c} Institute, P.O.Box 180, HR-10\,002 Zagreb, Croatia}
\address{$^2$ SUBATECH, EMN-IN2P3/CNRS-Universit\'e de Nantes, P.O.Box 20722, F-44\,307 Nantes, France}

\ead{basrak@irb.hr}

\begin{abstract}
Semiclassical transport simulation of nucleus-nucleus
collisions for the range of incident energy from about the
Fermi energy up to a few hundred MeV per nucleon evidences
that the maximal excitation energy put into a nuclear system
during the early compact stage of heavy-ion reaction is a
constant fraction of the center-of-mass available energy of
the system.
Analysis of experimental data without presuming reaction
mechanism dominating the collision process on the best
corroborate the found constancy of energy partition in
central heavy-ion reactions.
\end{abstract}

\section{Introduction}
Transformation of the entrance channel longitudinal motion
into internal degrees of freedom in the course of heavy-ion
reaction (HIR) is still awaiting for a satisfactory answer.
With increasing incident energy $E_{\rm in}$, in particular
for central
collisions, the reaction mechanism evolves from a slow,
essentially mean-field--transformation fusion and fusion-like
processes to a much faster and considerably more violent
reaction mechanism dominated by elementary nucleon-nucleon
(NN) collisions.
In fusion the entire available energy of the reaction is
deposited via thermal excitation, whereas at higher energy
a considerable fraction of the available energy is deposited
into system via compression.
By increasing $E_{\rm in}$ one expects that only a
fraction of the available energy is effectively deposited into
the reaction system and becomes dissipated during the reaction
course.
It is commonly admitted that this fraction should monotonically
decrease with the increase of $E_{\rm in}$.

At $E_{\rm in}$ from about the Fermi energy $E_{\rm F}$ to
about 100$A$ MeV, the energy transformation is determined by
those processes which govern heating and compression of a
reacting system.
In this energy range the time scales involved are rather
short and are of the order of time which reaction
partners need to bypass each other\ \cite{dst01,bas80}.
Two entrance channel factors play a central role in the
determination of the dominant reaction mechanism:
projectile energy per nucleon $E_{\rm in}$ and reaction
geometry (impact parameter and system mass asymmetry).
Consequently, the course of a HIR is "decided" in the
very first instances of a collision\ \cite{bon06r,eud97}.
In central, the most violent collisions the largest fraction
of the entrance channel energy is converted into internal
degrees of freedom.
Thus, the central collisions are of our greatest interest.

We have shown theoretically that an intermediate energy HIR
follows a two-stage scenario, a prompt first compact-stage
and a second after-breakup one\ \cite{eud97}.
The emission pattern of central collisions is characterized
by a copious and prompt dynamical emission occurring during
the compact and prior-to-scission reaction phase\
\cite{eud97,had99,eud00}.
This is the main system-cooling component and the amount
of deposited energy into the compact system linearly
increases with the projectile energy\ \cite{nov05}.
These results corroborate conclusion that global
characteristics of HIR exit channel are determined in the
first prompt reaction stage underlying the interest in
studying the first instances of nuclear collisions.

\begin{table}[b]
\begin{center}
\caption{\label{t-lv}Systems and energies studied for central
collisions.}
\centering
\begin{tabular}{@{}*{7}{l}}
\br
System&Incident energy ($A$ MeV)\\
\mr
$^{40}$Ar+$^{27}$Al & 25, 41, 53, 65, 77, 99\\
$^{36}$Ar+$^{58}$Ni & 52, 74, 95\\
$^{40}$Ar+$^{107}$Ag & 20, 30, 40, 45, 50, 75, 100\\
$^{40}$Ar+$^{197}$Au & 50, 75, 100\\
\mr
$^{36}$Ar+$^{36}$Ar & 32, 40, 52, 74\\
$^{58}$Ni+$^{58}$Ni & 52, 74, 90\\
$^{129}$Xe+$^{120}$Sn & 25, 32, 39, 45, 50, 75, 100\\
$^{197}$Au+$^{197}$Au & 20, 30, 40, 60, 80, 100\\
\br
\end{tabular}
\end{center}
\end{table}

In this work we theoretically examine how much of the system
energy may be temporarily stocked into the reaction system
in the form of excitation energy as a function of 
$E_{\rm in}$, system size $A_{\rm sys}$ and system mass
asymmetry.
Four mass symmetric and four mass asymmetric central reactions
were studied at several energies (see Tab.\ \ref{t-lv} for a
review).
Comparison with the pertinent results deduced from HIR
experiments is presented too.

\section{Theoretical approach}
Simulation was carried out within a semiclassical microscopic
transport approach of Boltzmann's type using the Landau-Vlasov
(LV) model\ \cite{rem-seb}.
The highly nonlinear LV equation
\begin{equation}
\frac{\partial f}{\partial t} + \{f,H\} = I_{\rm coll}(f)
\label{LVeq}
\end{equation}
\noindent
is solved by the test-particle method.
$f({\bf r},{\bf p};t)$ is the one-body density distribution
function describing the spatio-temporal evolution of the
system governed by the effective Hamiltonian $H$ consisting
of the self-consistent nuclear and Coulomb fields.
The D1-G1 momentum-dependent interaction due to Gogny
(the incompressibility module $K_{\infty}$=228 MeV and the
effective mass $m^*/m$=0.67)\ \cite{dec80} was used to
describe the nuclear mean-field potential.
$\{\;,\;\}$ stands for the Poisson brackets and
$I_{\rm coll}$ is the collision integral.
The effects of the Pauli-suppressed two-body residual NN
collisions are treated on average in
the Uehling-Uhlenbeck approximation taking the isospin- and
energy-dependent free-scattering value for the NN cross
section $\sigma_{\rm NN}$.
Such an approach is very successful in reproducing a variety
of global experimental dynamical observables because they are
adequately described by the time evolution of the one-body
density.
Thus, the LV model is especially appropriate for describing
the early stages of HIR, when the system is hot and compressed.

The observable studied is the thermal component (heat), i.e.\
one of the two main intrinsic-energy deposition components of
the early-reaction-stage energy transformation.
Heat is stocked into the compact system predominantly by NN
collisions which occurs in the overlap zone.
This is corroborated by the fact that the Pauli principle
greatly favors NN collisions involving one nucleon from the
target and one from the projectile.
In the most of cases under study the time is too short for
the full relaxation of the pressure tensor and establishment
of a global equilibrium in momentum space.
Therefore, it is more correct to name this component the
excitation energy $E_{\rm x}$.
Detailed definition of the transformation of the (system)
available energy $E_{\rm avail}^{\rm c.m.}$ into intrinsic
and collective degrees of freedom may be found elsewhere\
\cite{nov05,abg94,mota92}.
$E_{\rm avail}^{\rm c.m.}$ is defined as the center-of-mass
system energy per nucleon
$E_{\rm avail}^{\rm c.m.}=\frac{E_{\rm P}}{A_{\rm P}}
\frac{A_{\rm P} A_{\rm T}}{(A_{\rm P} + A_{\rm T})^2}$,
where $E_{\rm P}/A_{\rm P}=E_{\rm in}$ and $A_{\rm P}
(A_{\rm T})$ is the projectile (target) number of nucleons.

\section{Simulation results}
\begin{figure}[th]
\includegraphics[width=17pc]{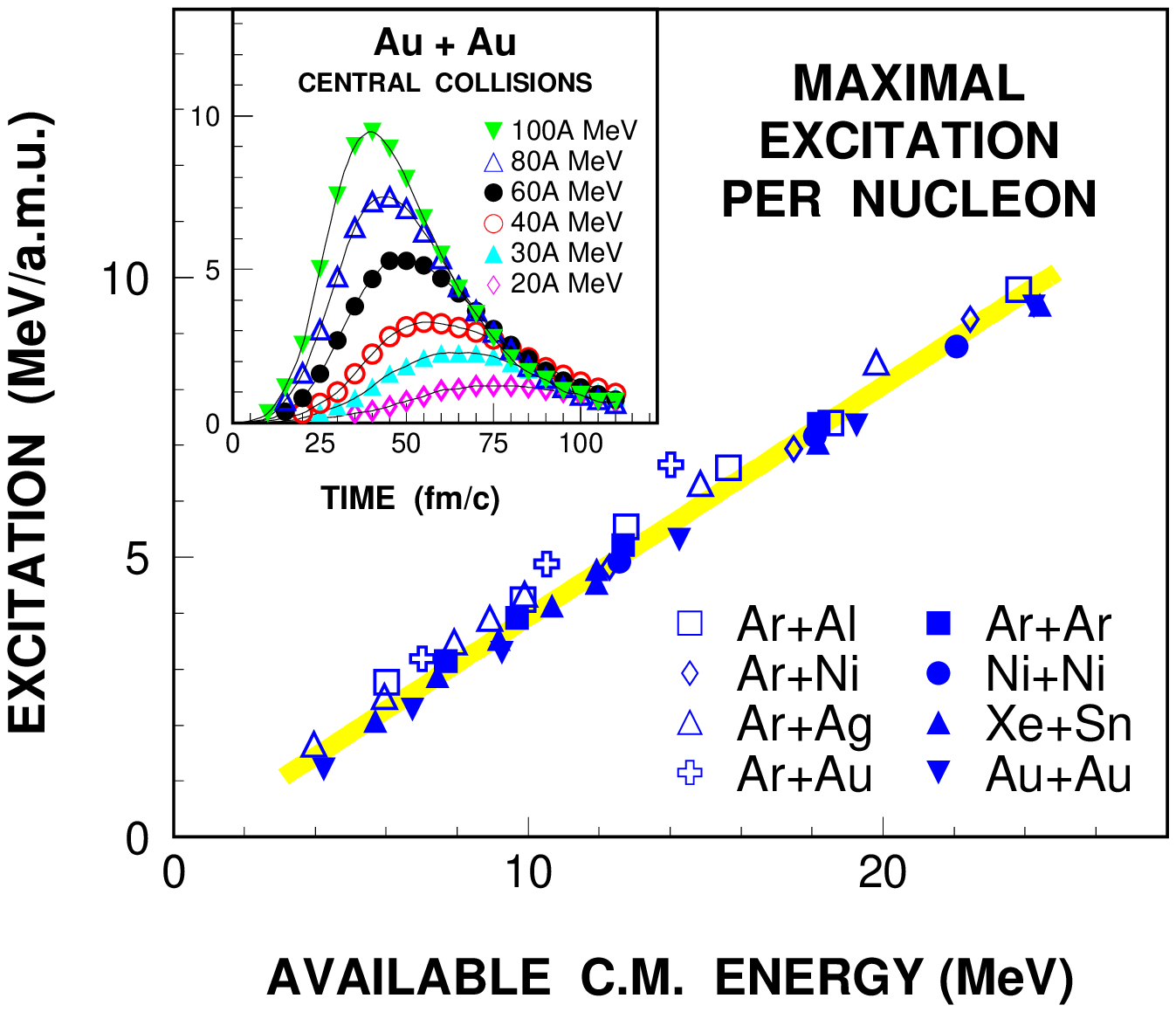}\hspace{1.5pc}%
\begin{minipage}[b]{19.4pc}\caption{\label{exc-lv}(Color online.)
Simulation results of the thermal excitation energy per
nucleon $E_{\rm x}/A$ for central collisions.

\noindent
\textit{Inset}\,:
Time evolution of $E_{\rm x}/A$ for the Au+Au reaction at
indicated energies.
At each time step considered are particles that are bound
in large fragments, in fact the early compact system.

\noindent
\textit{Main figure}\,:
Excitation maxima $(E_{\rm x}/A)_{\rm max}$ as a function of
system available energy $E_{\rm avail}^{\rm c.m.}$ for mass
asymmetric (open symbols) and mass symmetric (filled symbols)
systems studied.
The thick grey line is due to the best linear fit to all data
points.}
\end{minipage}
\end{figure}

As an example of the time evolution of excitation energy per
nucleon the inset of Fig.\ \ref{exc-lv} shows $E_{\rm x}/A$
for the Au+Au reaction at six energies studied.
Within a laps of time of merely 40--75 fm/$c$ after the
contact of colliding nuclei occurring at 0 fm/$c$ the 
excitation energy per nucleon $E_{\rm x}/A$ reaches a maximum
and then its value decreases almost as rapidly as it
increased.
As expected, the maxima are reached earlier and their height
increases and width decreases with increasing $E_{\rm in}$.
The regular and nearly symmetric rise and decrease of
$E_{\rm x}/A$ with the reaction time is a common behavior for
all reactions studied.
The observed regularity suggests that maxima of $E_{\rm x}/A$
are proportional to the total energy deposited during HIR.

We are examining the maximal energy that may be dissipated
in HIR.
Thus, we take the maxima of $E_{\rm x}/A$ which we denote by
$(E_{\rm x}/A)_{\rm max}$.
The value of $(E_{\rm x}/A)_{\rm max}$ can readily and
accurately be extracted from the simulation results.
Figure \ref{exc-lv} depicts how these maxima depends on
$E_{\rm avail}^{\rm c.m.}$ for all studied HIR.
Abscissa value is shifted for the threshold, the Coulomb
barrier energy.
With this correction the linear fit over all data points
crosses abscissa axis closer to the origin of the graph.
All data points lie very close to the fit line.
One is facing a peculiar universal linear rise which is
independent of $A_{\rm sys}$ and mass asymmetry in the full
and a rather large span of $E_{\rm in}$ covered in this study.
The linear dependence of $(E_{\rm x}/A)_{\rm max}$ on
$E_{\rm in}$ is, of course, present for each individual
system studied.
Specifying abscissae in $E_{\rm avail}^{\rm c.m.}$ rather
than in $E_{\rm in}$ merely expresses the mass asymmetric
systems on the same footing with those which are mass
symmetric.

\begin{figure}[th]
\includegraphics[width=14pc]{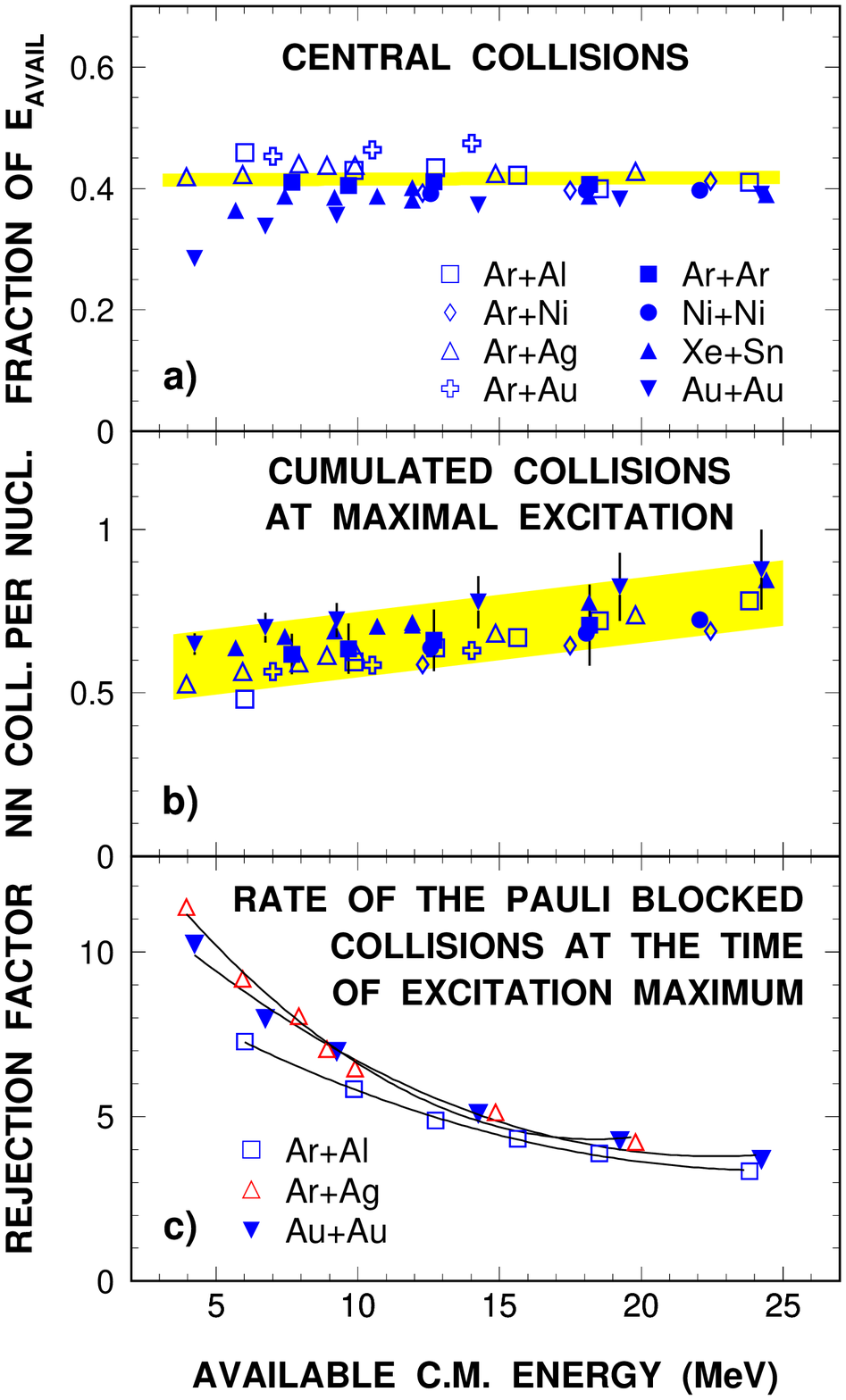}\hspace{1.5pc}%
\begin{minipage}[b]{22.4pc}\caption{(Color online.)
\textit{Top}\,:
Ratio of the excitation energy and the corresponding
$E_{\rm avail}^{\rm c.m.}$ as a function of this same
available energy $E_{\rm avail}^{\rm c.m.}$ for the
simulation results of Fig.\ \protect\ref{exc-lv}.

\textit{Middle}\,:
Cumulated average number of NN collisions per nucleon at the
time when $(E_{\rm x}/A)_{\rm max}$ is reached.

\textit{Bottom}\,:
The average value of the ratio of the potential to the
effectively realized NN collisions per nucleon at the time
when $(E_{\rm x}/A)_{\rm max}$ is reached.
The curves are due to a parabolic fit and are merely intended
to guide the eye.
\label{exc-r_lv} }
\end{minipage}
\end{figure}

A universal linear dependence of $(E_{\rm x}/A)_{\rm max}$
on $E_{\rm avail}^{\rm c.m.}$ as well as its nearly exact
crossing of the origin in Fig.\ \ref{exc-lv} has an
important and remarkable consequence:
Expressing the value of maximal excitation in percentage of
the system available energy one obtains that the relative
fraction of $(E_{\rm x}/A)_{\rm max}$ in
$E_{\rm avail}^{\rm c.m.}$ has an almost constant value
except for symmetric systems and $E_{\rm in}\!<\!E_{\rm F}$.
This departure from the constancy occurs because when
$E_{\rm in}$ decreases below $E_{\rm F}$\footnote{%
For mass symmetric systems $E_{\rm F}$ corresponds to
$E_{\rm avail}^{\rm c.m.}\!\approx\,$8$A$ MeV.}
the value of the maximum $(E_{\rm x}/A)_{\rm max}$ decreases
faster than $E_{\rm avail}^{\rm c.m.}$ itself is decreasing.
This is a consequence of an ever slower and slower the early
compact system energy transformation as $E_{\rm in}$ decreases
with an ever more broadened maximum (cf.\ inset in Fig.\
\ref{exc-lv}). 
Therefore, at these lower $E_{\rm in}$ the maximum
$(E_{\rm x}/A)_{\rm max}$ is no more proportional on the
same manner to the total energy deposited in HIR as for
$E_{\rm in}\geq E_{\rm F}$:
These $(E_{\rm x}/A)_{\rm max}$ cannot be compared with an
experimental $E_{\rm x}/A$ of fusion reaction, i.e.\ of
adiabatic-like processes.
With this restriction in mind, from Fig.\ \ref{exc-r_lv}a) one
infers that share of
$E_{\rm x}/A$ in $E_{\rm avail}^{\rm c.m.}$ weekly depends on
either reaction system or incident energy $E_{\rm in}$ and
amounts 0.39$\pm$0.03 of $E_{\rm avail}^{\rm c.m.}$.
In other words, during the early energy transformation in
HIR the maximal excitation energy that may be deposited in
the system is a constant which amounts about 40\,\% of the
system available energy.
As already discussed, this energy dissipation is chiefly
owing to NN collisions occurring in the reaction overlap zone.
At the time when $(E_{\rm x}/A)_{\rm max}$ takes place the
average number of NN collisions per nucleon bound in the
compact system for each individual system increases linearly
and slowly with $E_{\rm in}$
(cf.\ in Fig.\ \ref{exc-r_lv}b)), whereas Pauli suppression
of NN collisions approximately quadratically decreases with
$E_{\rm in}$ (cf.\ in Fig.\ \ref{exc-r_lv}c)).
In absolute values the above behaviors are within 20\,\% the
same for various systems in the energy range studied.
With the increase of $E_{\rm in}$, the energy transferred by
an individual NN collision and deposited within the overlap
zone increases on the average on such a way to remain
proportional to $E_{\rm avail}^{\rm c.m.}$.
The observed behavior of energy deposit in HIR is a
consequence of these facts.
Let us underline that this constancy of the
maximum-of-excitation-energy share in available energy is
evidenced in the fairly broad range of $E_{\rm in}$
(quotient of the highest and the lowest
$E_{\rm avail}^{\rm c.m.}$ covered in the simulation is
$\sim\,$9) and it is nearly independent of system size
(studied is the range of
$\sim\,$60$\,\leq\!A_{\rm sys}\!\leq\,\sim\,$400
nucleons) and mass asymmetry ($A_{\rm P}\!:\!A_{\rm T}$
is varied between 1:1 and 1:5).

\section{Comparison with experimental results}
An important question is whether the existing central HIR
experimental data support our simulation results.
Most of the energy put into the system during the early
reaction phase is released by the emission of particles and
light and intermediate mass fragments owing to the thermal
excitation component $E_{\rm x}$.
At energies below 100$A$ MeV the compression-decompression
process contributes a little in the total (kinetic) energy
dissipation in HIR\ \cite{poc97}.
At the instant at which the maximum $(E_{\rm x}/A)_{\rm max}$
is reached a negligible emission occurs.
At energies of our interest it amounts at most 3--5\,\% of
the total system mass\ \cite{nov05}.
Thus, conjunction of the $(E_{\rm x}/A)_{\rm max}$ with the
total (kinetic) energy released in HIR seems to be a natural
assumption.
One must keep in mind, however, that a simulation maximum
is reached prior to although very close (of the order of
$\sim$5--10 fm/$c$)
to the time at which the total momentum distribution becomes
locally spherical, i.e., the instant at which the local
equilibrium has been reached in each part of the compact
subsystem of bound particles\ \cite{abg94}.
Nevertheless, the system is far from a global equilibrium\
\cite{nov05} and comparison with experimental $E_{\rm x}/A$
is not straightforward.
In addition, one must bear in mind that one should limit the
comparison to
general trend of experimental data, i.e. to the possible
constancy of $(E_{\rm x}/A)/E_{\rm avail}^{\rm c.m.}$ as a
function of $E_{\rm avail}^{\rm c.m.}$ without seeking to
reproduce the simulation absolute value.
The maximal excitation share of $(E_{\rm x}/A)_{\rm max}$ in
$E_{\rm avail}^{\rm c.m.}$ of 40\,\% is reached during the
very first reaction phase and if the same value would be
extracted from experimental data that could not but be a
fortuitous result.
Indeed, experimental data is registered at an infinite time.
Hence, it reflects an integral of the full reaction history.
Anyway, the simulation maxima
$(E_{\rm x}/A)_{\rm max}$ should be compared with either
the maximal value of $E_{\rm x}/A$ obtained in an experiment
or with the most probable value of $E_{\rm x}/A$ depending on
the nature of the distribution.

\begin{figure}[th]
\includegraphics[width=18pc]{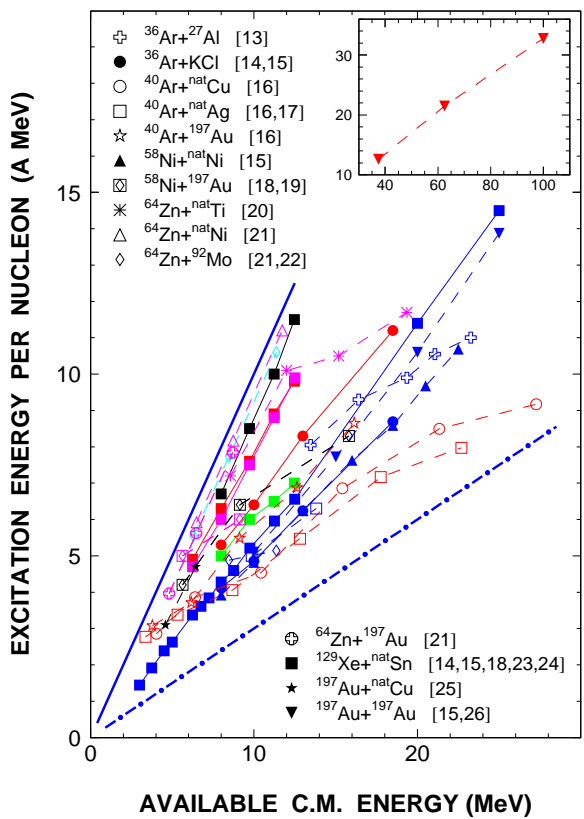}\hspace{1.5pc}%
\begin{minipage}[b]{18.4pc}\caption{(Color online.)
Experimentally evaluated total excitation energy per nucleon
or total dissipated energy per nucleon as a function of
system available energy.
Each reaction is represented by its own symbol and by color
are distinguished different analyses of the same reaction.
The thick full line corresponds to the $E_{\rm avail}^{\rm c.m.}$
and displays the upper energy limit which may be reached in
HIR while the thick dash-dotted line depicts 30\,\% of this
limit.
The only data on total $E_{\rm x}/A$ above 100$A$ MeV are
for the Au+Au reaction at $E_{\rm in}$=150$A$, 250$A$ and
400$A$ MeV\ \protect\cite{rei97}.
They are shown in the inset.
The axes aspect ratio of both the inset and the main diagram
is the same so that the slope in both is the same.
\label{exc-exp} }
\end{minipage}
\end{figure}

Figure \ref{exc-exp} displays a collection of experimental
data on $E_{\rm x}/A$ and total energy dissipated in central
HIR published in periodics during the last two decades\
\cite{pet95,met00,leh00t,rul00,vie94,bor04,bel02,ste96,wan05b,wan05,len07,bon10,dag03,rei97}.
Because energy dependence is crucial for our comparison
from the figure are dropped all single-energy results.
Each reaction system is depicted by its symbol while the
different measurements of the same system are distinguished
by color (on line).
To avoid of entirely spoiling the figure the error bars,
typically of 5--15\,\%, are not displayed.
To guide the eye, points belonging to the same system and the
same analysis are connected and they mostly display
close-to-linear dependence on $E_{\rm avail}^{\rm c.m.}$.
Unlike the simulation result on $(E_{\rm x}/A)_{\rm max}$
the experimental data points span a large domain of the
$E_{\rm x}/A$ vs.\ $E_{\rm avail}^{\rm c.m.}$ plane:
The extracted excitations per nucleon lie between one third
and almost the full accessible system energy
$E_{\rm avail}^{\rm c.m.}$.
One may speculate that the
different approaches used in extracting from experiments the
pertinent information on the global energy deposition in HIR
might be at the origin of these much more scattered results.
Indeed, in a HIR experiment one does not have a direct access
to the excitation energies involved.
To obtain $E_{\rm x}/A$ one needs to reconstruct from
detected reaction products the total excitation $E_{\rm x}$
of an assumed primary emission source but also the source
mass $A$.
There is an evident difficulty to restore the break-up stage
using exclusively asymptotic experimental information which
is further obscured by an important role played by primary
fragments internal excitation causing the in-flight emission.
To overcome these uncontrolled issues one has to resort to
certain more or less justified physical assumptions or/and
to use theoretical predictions as a guide for data analysis.
Anyhow, data analyzed on a same footing seems to fall into
much narrower zones of the $E_{\rm avail}^{\rm c.m.}$ vs.\
$E_{\rm x}/A$ plane.

\begin{figure}[th]
\includegraphics[width=18pc]{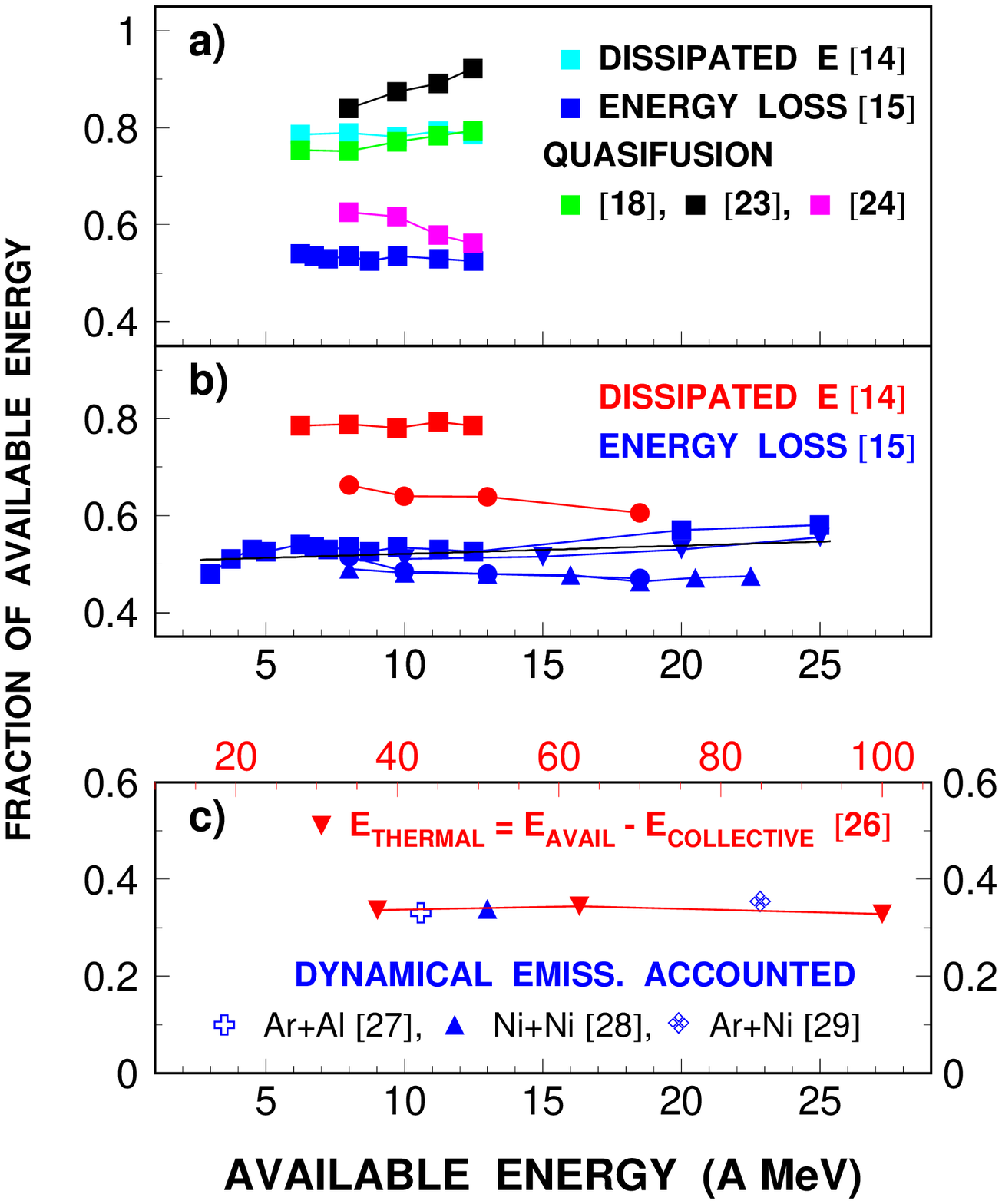}\hspace{1.5pc}%
\begin{minipage}[b]{18.4pc}\caption{(Color online.)
Ratio of the excitation energy and the corresponding
$E_{\rm avail}^{\rm c.m.}$ as a function of this same
available energy $E_{\rm avail}^{\rm c.m.}$.
Symbols used to distinguish different systems are the
same as in Fig.\ \protect\ref{exc-exp}.

\textit{Top}\,:
Five different analysis of the Xe+Sn reaction for
25$A\!\leq\!E_{\rm in}\!\leq\,$50$A$ MeV.

\textit{Middle}\,:
Ratio values reported in the analyses based on the pure
kinematical considerations.

\textit{Bottom}\,:
Ratio values reported in analyses which thoroughly
accounted for the pre-equilibrium emission component as well
as the results on the total thermal energy reported above
100$A$ MeV and for which the abscissae labels above the
graph frame are relative to.
\label{exc-r_exp} }
\end{minipage}
\end{figure}

Linear dependence of $E_{\rm x}/A$ on $E_{\rm avail}^{\rm c.m.}$
is not sufficient to obtain a constancy of its fraction in
available energy.
For this constancy the line passing through data points
should also pass close to the origin of the
$E_{\rm avail}^{\rm c.m.}$ vs.\ $E_{\rm x}/A$ plane.
As an example in Fig.\ \protect\ref{exc-r_exp}a) are shown
results for the Xe+Sn system which have been extensively
studied by the INDRA collaboration.
Displayed are five analyses of apparently the same data set
for $E_{\rm in}$ between 25$A$ and 50$A$ MeV\
\cite{met00,leh00t,bor04,len07,bon10}.
Each analysis have used its own approach in selecting data
by centrality and its own philosophy reagrding the presumed
dominant reaction mechanism used to extract the total
excitation $E_{\rm x}$ and the primary source mass $A$.
Reported $E_{\rm x}/A$ differ substantially among them.
The absolute value at the same $E_{\rm in}$ differs up to
80\,\%.
In addition, some of presumed single-source analyses
display a rising fraction of $E_{\rm x}/A$ in
$E_{\rm avail}^{\rm c.m.}$ as $E_{\rm in}$ increases\
\cite{bor04,len07}, other falling fraction as $E_{\rm in}$
increases\ \cite{bon10}, whereas the most probable dissipated
energy\ \cite{met00} and the total energy loss\ \cite{leh00t}
displays a weak if any dependence on $E_{\rm in}$.
One may argue that various selections of central events may
reflect different physics and, thus, have a different
$E_{\rm x}/A$.
One must admit that reported differences, including rising,
steady and falling behavior seems to be too big.
This indicates how delicate these analyses are as well as
that at least some of them may contain incorrect step(s) in
the extraction of the reported values.

Dissipated energy and total energy loss are the analyses
inspired by the
kinematical arguments and do not require presumption on the
dominant reaction mechanism.
Their drawback is in their applicability to the
mass-symmetric systems only.
Figure \protect\ref{exc-r_exp}b) displays results for all
systems studied by these two approaches in a fairly broad
range of $E_{\rm in}$.
The total energy loss within the error bars gives the same
constant value for all four systems studied.
These results are rather weekly depending on $E_{\rm in}$
and may be considered constant.
Another example of cases with the constant fraction of
$E_{\rm x}/A$ in $E_{\rm avail}^{\rm c.m.}$ is shown in Fig.\
\protect\ref{exc-r_exp}c).
Displayed are three single-energy
studies that carefully accounted for the copious midrapidity
emission which occurs during the compact and prior-to-scission
reaction phase discussed earlier\ \cite{lan01,the05,dor00}
as well as the only $E_{\rm x}/A$ result reported so far
above 100$A$ MeV.
Within blast model extracted is the total thermal energy for
the Au+Au reaction from 150$A$ to 400$A$ MeV\ \cite{rei97}.
The abscissae labels above the graph frame are relative to
the Au+Au reaction.
These Au+Au data have recently been revised\ \cite{rei10}
but a strict linearity of the studied ratio as a function of
$E_{\rm in}$ did not change so that the value of our
fraction should merely be slightly increased.

\section{Conclusions}
In conclusion, a semiclassical transport model study of the
early reaction phase of central heavy-ion collisions at
intermediate energies has been carried out for a variety of
system masses, mass asymmetries, and energies below 100$A$
MeV.
It has been found that the maxima of the excitation energy
$E_{\rm x}$
deposited at this early reaction stage into the reaction
system represents a constant fraction of about 40\,\% of the
total center-of-mass available energy of the system
$E_{\rm avail}^{\rm c.m.}$.
In heavy-ion experiments extracted total dissipated energy
per nucleon and total energy loss deduced on kinematical
arguments display a similar constancy of their share in the
system available energy.
A similar result may be found in total excitation energy
extracted from experimental observations under condition
that the pre-equilibrium emission is properly accounted for.
These results indicate that the stopping power of nuclear
matter is significant even below the threshold of nucleon
excitation and that it does not change appreciably when
expressed in units of the center-of-mass available system
energy over a wide range of incident energies.

\end{document}